\documentclass[twocolumn,journal]{IEEEtran}
\usepackage[T1]{fontenc}
\usepackage{amsmath}
\usepackage{amssymb}
\usepackage{graphicx}
\usepackage{bbm}
\usepackage{float}
\usepackage{booktabs}
\usepackage{nccbbb}
\usepackage{booktabs,multirow}
\usepackage{makecell}
\usepackage[normalem]{ulem}
\usepackage[unicode=true,
 bookmarks=true,bookmarksnumbered=true,bookmarksopen=true,bookmarksopenlevel=1,
 breaklinks=false,pdfborder={0 0 0},pdfborderstyle={},backref=false,colorlinks=false]
 {hyperref}
\hypersetup{pdftitle={Self-Supervised Knowledge-Driven Deep Learning for 3D Magnetic Inversion},
 pdfauthor={Yinshuo Li},
 pdfpagelayout=OneColumn, pdfnewwindow=true, pdfstartview=XYZ, plainpages=false}

\makeatletter
 \let\oldforeign@language\foreign@language
 \DeclareRobustCommand{\foreign@language}[1]{%
   \lowercase{\oldforeign@language{#1}}}

\usepackage[caption=false,font=footnotesize]{subfig}

\usepackage{cite}
\usepackage{color}

\makeatother
\usepackage{amsmath,bm}

\begin{document}

%
\title{Self-Supervised Knowledge-Driven Deep Learning \\ for 3D Magnetic Inversion}

\author{Yinshuo~Li,
        ~Zhuo~Jia,
        ~Wenkai~Lu$^{*}$,~\IEEEmembership{Member,~IEEE},
        ~Cao~Song

\thanks{
This research is financially supported by the National Key R $\&$ D Program of China (Grant No. 2018YFA0702501) and the NSFC (Grant No. 41974126).
}

\thanks{Y. Li and W. Lu are with Beijing National Research Center for Information Science and Technology (BNRist), Department of Automation, Tsinghua University, Beijing, P.R.China
(e-mail: ys-li22@mails.tsinghua.edu.cn, lwkmf@mail.tsinghua.edu.cn).
}
\thanks{Z. Jia is from the School of Civil Engineering, Changsha University of Science and Technology, Changsha, China
(e-mail: jiazhuo@mail.tsinghua.edu.cn).
}

\thanks{$^{*}$ Correspondence Author: lwkmf@mail.tsinghua.edu.cn}
}

%
%

\markboth{}
{JIA \MakeLowercase{\emph{et al.}}: Self-Supervised Knowledge-Driven Deep Learning for 3D Magnetic Inversion}

\maketitle

\begin{abstract}
The magnetic inversion method is one of the non-destructive geophysical methods, which aims to estimate the subsurface susceptibility distribution from surface magnetic anomaly data. Recently, supervised deep learning methods have been widely utilized in lots of geophysical fields including magnetic inversion. However, these methods rely heavily on synthetic training data, whose performance is limited since the synthetic data is not independently and identically distributed with the field data. Thus, we proposed to realize magnetic inversion by self-supervised deep learning. The proposed self-supervised knowledge-driven 3D magnetic inversion method (SSKMI) learns on the target field data by a closed loop of the inversion and forward models. Given that the parameters of the forward model are preset, SSKMI can optimize the inversion model by minimizing the mean absolute error between observed and re-estimated surface magnetic anomalies. Besides, there is a knowledge-driven module in the proposed inversion model, which makes the deep learning method more explicable. Meanwhile, comparative experiments demonstrate that the knowledge-driven module can accelerate the training of the proposed method and achieve better results. Since magnetic inversion is an ill-pose task, SSKMI proposed to constrain the inversion model by a guideline in the auxiliary loop. The experimental results demonstrate that the proposed method is a reliable magnetic inversion method with outstanding performance.
\end{abstract}


\IEEEpeerreviewmaketitle

\section{Introduction}

\IEEEPARstart{T}he magnetic method is a crucial non-destructive method for geophysical exploration \cite{li19963}, which has been widely used in environment-friendly mineral and petroleum exploration \cite{dhiman2021nanostructured}. Besides, the 3D magnetic inversion hammer at obtaining 3D subsurface magnetic susceptibility distribution from 2D surface magnetic anomaly data, which is more practical than 2D magnetic inversion \cite{portniaguine20023}. Since it aims to estimate a 3D matrix from 2D data, 3D magnetic inversion is an ill-posed problem with a non-unique solution space \cite{chevalier2014monte}. Oldenburg et al. \cite{oldenburg2005inversion} have demonstrated that the properly designed inversion objective function is a prerequisite for the accurate estimation of complex magnetic anomaly. However, the magnetic anomaly is 2D surface data, which lacks depth resolution. According to the Gauss theorem, the 3D magnetic inversion suffers from serious multiplicity solutions inevitably since it aims at estimating 3D structure from 2D data. Many methods have been tried to relieve this problem and improve the practicability of magnetic inversion \cite{li19963, jia2023deep}.

Traditional 2D and 3D magnetic inversion methods of ground and aeromagnetic data suffer from insufficient vertical resolution, which leads to multiple solutions and uncertainty of interpretation \cite{jia2023deep}. Parker and Huestis \cite{parker1974inversion} point out that susceptibility anomalies at different depths in the same horizontal area can easily lead to similar surface magnetic anomalies, which is known as the volume effect. Besides, according to magnetic Coulomb’s law \cite{frohlich1986stability}, the influence of magnetic anomaly decays rapidly with the increase of distance. Thus, the surface magnetic anomaly data is sensitive to the near-surface magnetic susceptibility \cite{foote1996relationship}, which leads to the loss of sensitivity to the deeper. Since the near-surface magnetic susceptibility takes the lead of the surface magnetic anomaly, 3D magnetic inversion suffers from the ``skin effect’’ \cite{wheeler1942formulas}. Specifically, once the near-surface magnetic susceptibility is well estimated, the re-estimated surface magnetic anomaly data will fit the target and the deeper magnetic susceptibility will be ignored. Since lots of magnetic methods have utilized constraints of regularization, the estimated deeper subsurface magnetic susceptibility tends to be zero or similar to the near-surface magnetic susceptibility \cite{uno2009modeling}.

In order to suppress the above shortage of inversion, researchers have employed depth weighting and other geological prior information into the objective function of traditional magnetic inversion \cite{lan20153d, guo20213d, rossi2015integrating}. Since the magnetic anomaly gradient data contains abundant geological information, joint inversion of magnetic anomaly and its gradient data has been utilized to improve the inversion results \cite{pilkington19973}. Besides, joint inversion of magnetic and other inversion methods, e.g., well log and gravity inversion, are widely used to obtain better inversion results \cite{meng20123, fedi19993, zhdanov2012generalized}. However, these inversion methods are highly dependent on the initial model and their stability is poor \cite{shamsipour20113d}. Specifically, these linear inversion methods without well-selected initial models often fall into local optimum \cite{zhang2021robust}. Meanwhile, useful prior information is difficult to be effectively applied to complex geological conditions by traditional magnetic inversion methods \cite{huang2020geological}.

\begin{figure*}[t]
\centering
\includegraphics[width=0.7\textwidth]{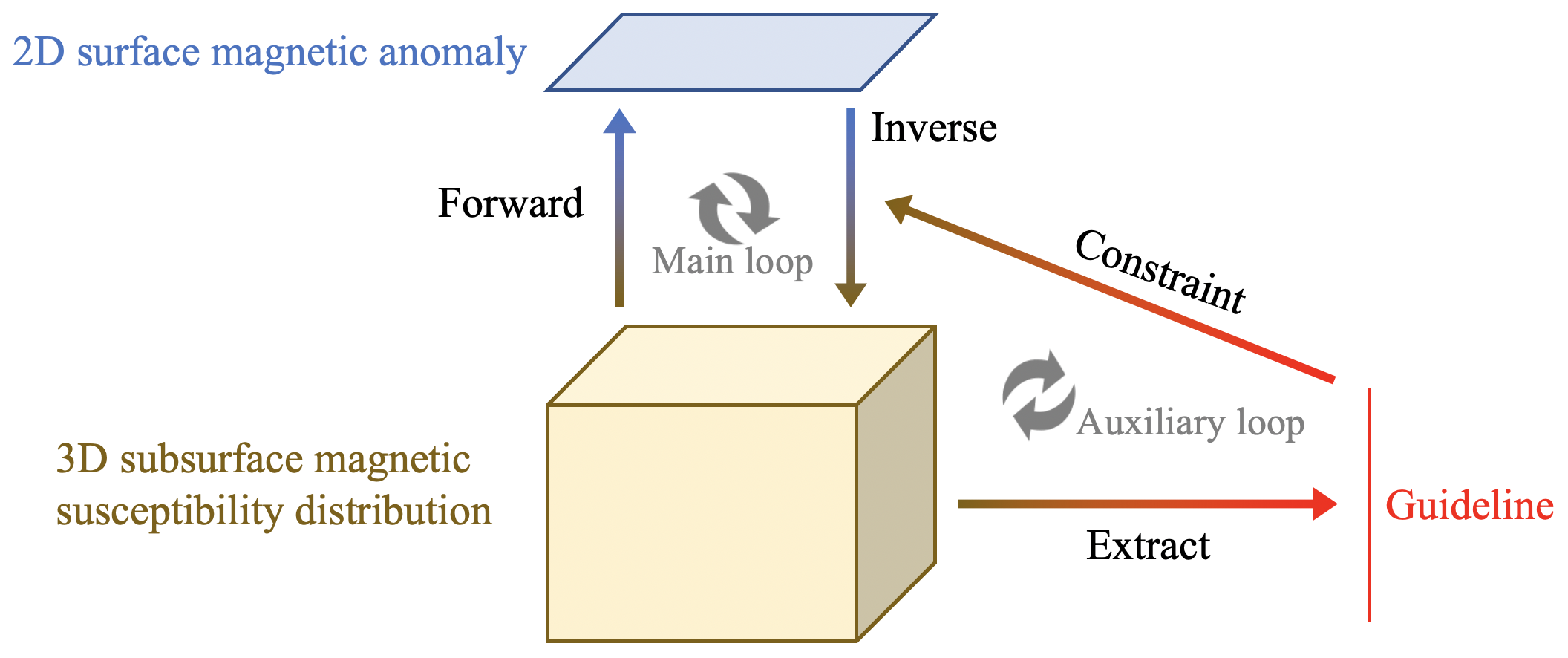}
\caption{Pipeline of the proposed method. The proposed method employs two closed loops to optimize the parameters of the inversion model.}
\label{fig_pipeline}
\end{figure*}

Benefiting from the development of computer hardware especially the graphics processing unit (GPU), the artificial neural network has achieved outstanding performance in computer vision \cite{ho2020denoising} and natural language processing \cite{vaswani2017attention}. Pixel-level restoration tasks in computer vision, e.g., super-resolution \cite{saharia2022image}, denoising \cite{kawar2022denoising}, and defog \cite{zhu2021remote}, are committed to restoring high-quality information from the observed images. Methods based on the convolutional neural network (CNN) for these tasks have achieved rapid development in recent years \cite{chen2021pre}. Since the target of image super-resolution is similar to that of physical property inversion, the model structures of super-resolution have reference significance on the geophysical inversion \cite{li2022self}. Lim et al. \cite{lim2017enhanced} introduced residual blocks (RBS) to image super-resolution and removed the batch normalization (BN) module to achieve better results in pixel-level restoration. Besides, Zhang et al. \cite{zhang2018residual} introduced residual dense blocks (RDBs) to pixel-level restoration tasks and achieved better results. Meanwhile, to enable a more powerful feature representation, Yang et al. \cite{yang2020learning} introduced cross-scale feature fusion to pixel-level restoration tasks. It has been proven that these CNN structures are effective, robust, and fast, while all of them rely on plenty of high-quality training data \cite{chen2021learning}.

Since the essence of an image is a matrix of pixels, geophysical data can be regarded as images and processed by CNN. Li et al. \cite{li2020multitask, li2021super} utilized multitask deep learning for super-resolution of the seismic velocity model, which improved the efficiency and effect of full waveform inversion. Besides, physical property inversion aims to estimate subsurface information from the observed surface data. Given that the targets of physical property inversion and pixel-level restoration tasks are similar, researchers have proposed to realize physical property inversion by deep learning (DL), especially CNN. Wang et al. \cite{wang2020well} and Zhang et al. \cite{zhang2021robust} realized constrained seismic inversion by closed-loop of CNN. Besides, Zhang et al. \cite{zhang2021deep} achieved a DL-based 3D gravity inversion method, which could obtain reliable gravity inversion results. It is exciting that Jia et al. \cite{jia2023deep} have demonstrated that the 3D magnetic inversion tasks can be solved by deep learning. However, these deep learning methods based on CNN rely heavily on high-quality training data and cannot adjust the inversion results according to prior information \cite{li2022self}.  Thus, the lack of high-quality training data is a major obstacle to applying DL technology in magnetic inversion \cite{scher2018toward}. Synthesizing training data based on expert knowledge is a feasible solution to solve the problem \cite{li2021super}. However, its effectiveness is limited since the synthetic data is not independently identically distributed with the field data.


To overcome the above drawback of DL-based inversion methods and obtain better estimate results on field magnetic data, we proposed a self-supervised knowledge-driven 3D magnetic inversion method (SSKMI). As shown in Fig. \ref{fig_pipeline}, SSKMI is based on closed loops of CNNs. Specifically, the proposed method builds two neural networks to realize inversion and forward process respectively. Besides, the parameters of the forward model are solved by magnetic Coulomb’s law. Thus, the parameters of the inversion model can be optimized by the closed loop between the forward model and the inversion model. Meanwhile, there is a design of a knowledge-driven module to accelerate the inversion model’s optimization and achieve better results. Furthermore, the proposed SSKMI employed a guideline to constrain the inversion model in the vertical direction. The main contributions of this work include the following aspects:

(1) The proposed SSKMI is a self-supervised magnetic inversion method, which does not require additional synthetic training data;

(2) Our method utilizes prior knowledge to reduce the uncertainty of magnetic inversion in-depth direction;

(3) The inversion model of SSKMI contains a knowledge-based module, which improved the interpretability of the proposed DL-based magnetic inversion method.

(4) The proposed method is a robust and efficient magnetic inversion method, which achieves state-of-the-art performance.

The rest of this paper is organized as follows: In Section \ref{Theory}, we describe the details of the proposed self-supervised magnetic inversion method, including the structure of the forward and inversion model, the two closed loops, the loss function, and the optimization strategy. Then, the experimental setups and results are demonstrated and other magnetic inversion methods are compared in Section \ref{Experiments}. Finally, our conclusion is drawn based on the conducted experiments in Section \ref{Conclusion}.

\section{Theory}\label{Theory}

\begin{figure*}[t]
\centering
\includegraphics[width=0.8\textwidth]{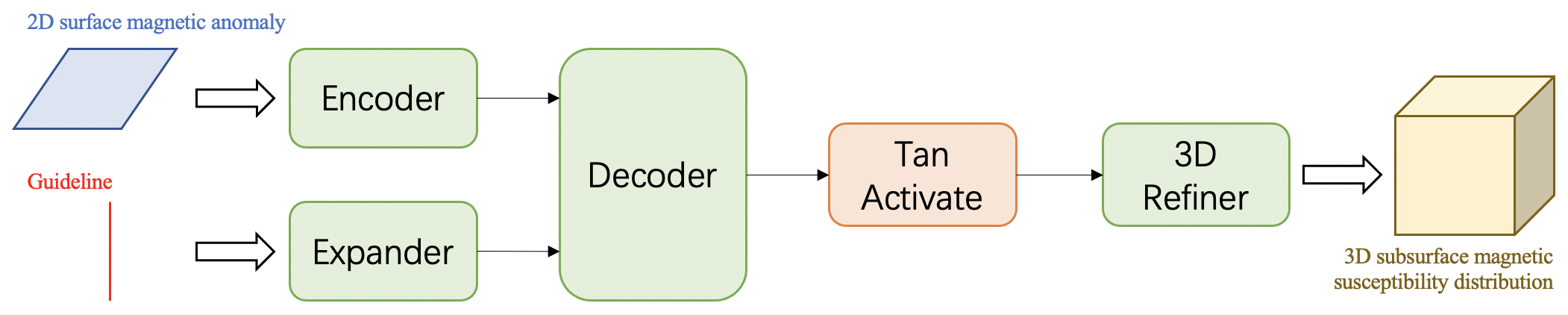}
\caption{Inversion model of the proposed SSKMI.}
\label{fig_inversion}
\end{figure*}

The proposed method aims at estimating 3D subsurface magnetic susceptibility distribution from the given 2D surface magnetic anomaly and 1D guideline. As shown in Fig. \ref{fig_pipeline}, the proposed SSKMI contains two closed loops. The main loop is built by two neural networks, they achieve the forward process and inversion process, respectively. The auxiliary loop of 3D subsurface magnetic susceptibility distribution and guideline is utilized to constrain the inversion result in the depth direction. Besides, the inversion model is the optimized model for 2D-to-3D inversion and contains a knowledge-driven module to improve the interpretability of DL-based inversion. The details of SSKMI are described in this section.

\subsection{Forward Model Realized by the Neural Network.}

The forward model is based on Coulomb’s law (\cite{shabad2007modified}), which is realized by the neural network in this paper. First, the 3D subsurface magnetic susceptibility and the 2D surface magnetic anomaly were differentiated separately. According to Coulomb’s law, for the surface observer, the magnetism of the cube region can be expressed as follows:
\begin{equation}
\begin{split}
Z=&-M\frac{\mu_{0}}{4\pi}\left(\cos\alpha_{s}\left |\left | \left |\ln[r^2+(y-\eta )] \right |_{\xi_1}^{\xi_2}\right |_{\eta_1}^{\eta_2}\right |_{\zeta_1}^{\zeta _2}\right) \\
&-M\frac{\mu_{0}}{4\pi}\left(\cos\beta_{s}\left |\left | \left |\ln[r^2+(x-\xi )] \right |_{\xi_1}^{\xi_2}\right |_{\eta_1}^{\eta _2}\right |_{\zeta_1}^{\zeta _2}\right) \\
&+M\frac{\mu_{0}}{4\pi}\left(\cos\gamma _{s}\left |\left | \left | \text{arctan}\frac{(x-\xi)(y-\eta)}{(z-\zeta)r} \right |_{\xi_1}^{\xi_2}\right |_{\eta_1}^{\eta _2}\right |_{\zeta_1}^{\zeta _2}\right),
\end{split}
\label{eq_law}
\end{equation}
where $\mu_0$ stands for vacuum susceptibility, while $M$ stands for the magnetization intensity of a subsurface cube region. Besides, $r=[(x-\xi)^2+(y-\eta)^2+(z-\zeta)^2]^{1/2}$ represents the distance between the surface observer and the cube region. The relative location of the subsurface cube region and the surface observer is described by $(x,y,z)$, where $z$ is the location in the depth direction. $\xi_1$ and $\xi_2$ stand for the minimum and maximum coordinates of cube region $j$ in the $x$ direction. Similarly, $\eta_1, \eta_2$ and $\zeta_1, \zeta_2$ represent the boundary in $y$ direction and $z$ direction, respectively. $\alpha_s$, $\beta_s$, and $\gamma_s$ are the angle of magnetic dip in the $x$, $y$, and $z$ directions. 

Because of the collected mag anomalies are polarized (\cite{li2003fast}), $\alpha_s$ and $\beta_s$ are set to $\pi/2$, while $\gamma_s=0$. Thus, Eq. \ref{eq_law} can be simplified as follows:
\begin{equation}
Z=M\frac{\mu_0}{4\pi}\left|\left|\left|arc\tan\frac{(x-\xi)(y-\eta)}{(z-\zeta)r}\right|_{\xi_1}^{\xi_2}\right|_{\eta_1}^{\eta_2}\right|_{\zeta_1}^{\zeta_2},
\label{eq_law_sim}
\end{equation}
Then multiple observation points on the 2D surface are collected to form a matrix of observation points to obtain surface magnetic anomaly data. The forward process of magnetic in Eq. \ref{eq_law_sim} can be expressed as follows:
\begin{equation}
Z=\mathcal{F}(M)=AM=\sum_i\sum_j a_{i,j}m_j,
\label{eq_forward}
\end{equation}
where $Z$ represents the surface magnetic anomaly, while $M$ stands for the 
subsurface magnetization intensity. Besides, $A$ stands for the forward matrix, which is the parameter of the forward neural network $\mathcal{F}$. The expression of the magnetic forward process is expressed as matrix multiplication in Eq. \ref{eq_forward}, which can be realized by a neural network. Meanwhile, if the area of observed magnetic anomaly is huge enough, a convolutional neural network is recommended to reduce computation.

\begin{figure*}[t]
\centering
\includegraphics[width=1\textwidth]{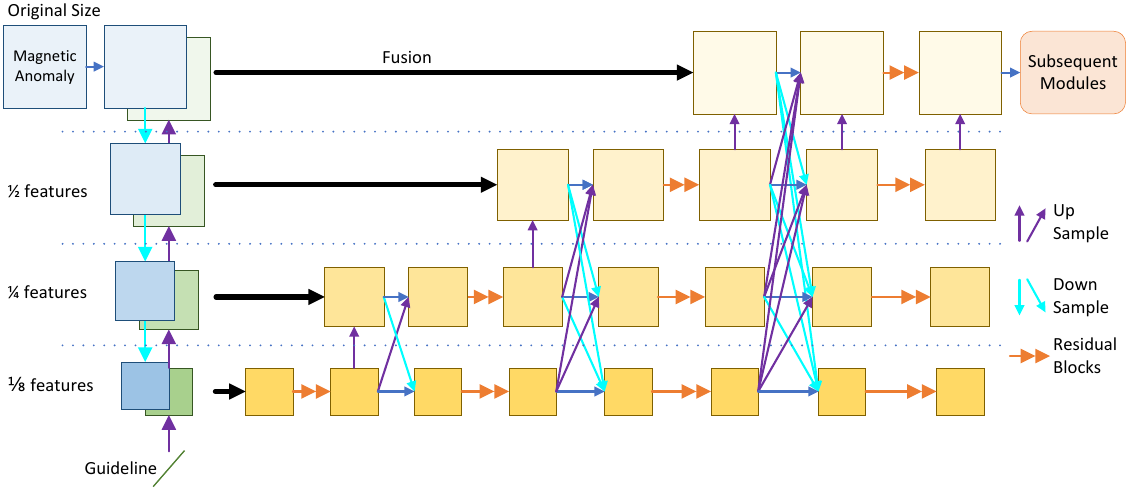}
\caption{2D feature extract of the proposed inversion model. This figure shows the partial structure of the inversion model, including the encoder, expander, and decoder. The subsequent modules compose of the knowledge-driven module, 2D-3D transform module, and 3D residual blocks.}
\label{fig_network}
\end{figure*}

\subsection{Explicable DL Method for Inversion.}

The proposed SSKMI obtains subsurface magnetic susceptibility distribution from the observed surface magnetic anomaly and the guideline by the inversion model. The guideline is utilized to constrain the uncertainty of inversion and improve the interpretability of the inversion model since magnetic inversion methods are troubled by the non-unique solution space. The expression of the proposed inversion model is as follows:
\begin{equation}
M=\mathcal{I}(Z,L;\theta),
\label{eq_forward_l}
\end{equation}
where $\theta$ stands for parameters of the inversion model $\mathcal{I}$, while $Z$ and $L$ represent the observed 2D surface magnetic anomaly and 1D guideline, respectively. Meanwhile, $M$ denotes the estimated 3D subsurface magnetic susceptibility distribution.

Fig. \ref{fig_inversion} shows that the proposed inversion model contains an encoder, an expander, a decoder, a 3D refiner, and a ``Tan Activate'' module. The encoder, decoder, and 3D refiner are based on 2D or 3D residual blocks (\cite{he2016deep}) without batch normalization (\cite{ioffe2015batch}), which has been proven to be applicable to pixel-to-pixel estimation (\cite{lim2017enhanced}). The encoder extracts feature maps from the observed magnetic anomaly, while the expander expands 2D features from the guideline. These two groups of features are fused and further extracted in the decoder. Finally, the output 2D features of the decoder are reshaped into 3D and put into the refiner to obtain the 3D subsurface magnetic susceptibility distribution.

\begin{figure}[t]
\centering
\includegraphics[width=1\linewidth]{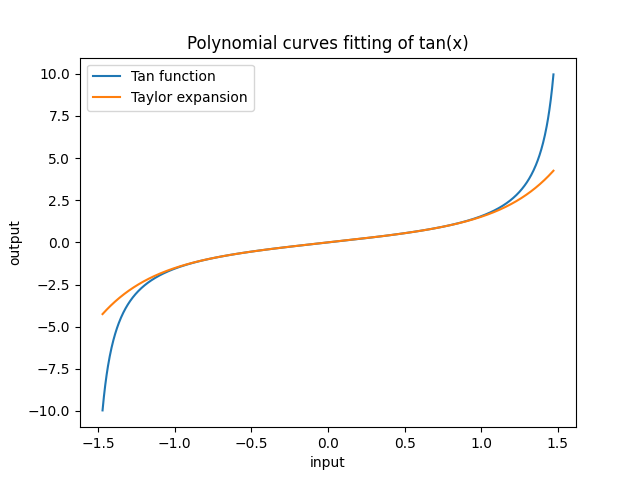}
\caption{Polynomial curves fitting of tan function by Taylor expansion.}
\label{fig_tan}
\end{figure}

It is necessary to note that the proposed inversion model employs a ``Tan Activate'' module as the inversion of ``arctan'' in Eq. \ref{eq_law_sim}, which makes the DL-based inversion model explicable. Besides, the ``Tan Activate'' module saves SSKMI from fitting ``tan'' by the convolution layers. The continuous input intervals of ``tan'' is from $-\pi/2$ to $\pi/2$, but the signal may go outside this range during neural network training. Thus, we proposed to utilize Taylor's expansion of the ``tan'' function in the inversion model. The Taylor expansion of the ``tan'' function is expressed as follows:
\begin{equation}
\begin{split}
    \tan x=&\sum^{\infty}_{n=0}\frac{U_{2n+1}x^{2n+1}}{(2n+1)!}\\
    =&\sum^{\infty}_{n=1}\frac{(-1)^{n-1}2^{2n}(2^{2n}-1)B_{2n}x^{2n-1}}{(2n)!}\\
    =&x+\frac{x^3}{3}+\frac{2x^5}{15}+\frac{17x^7}{315}+\cdots,
\end{split}
\label{eq_tan}
\end{equation}
in which $x\in(-\pi/2, \pi/2)$ is the domain of definition. Fig. \ref{fig_tan} shows the difference between the ``tan'' function and its Taylor expansion in the range of $(-\pi+0.05, \pi-0.05)$. As shown in Fig. \ref{fig_tan}, Taylor expansion $x+{x^3}/{3}+{2x^5}/{15}+{17x^7}/{315}$ is accurate polynomial fitting curve of ``tan'' function when $x\in(-1, 1)$. Besides, the valid input of Taylor expansion is $(-\infty, +\infty)$, which is more suitable for deep learning networks.

\begin{figure*}[t]
\centering
\noindent\includegraphics[width=0.9\linewidth]{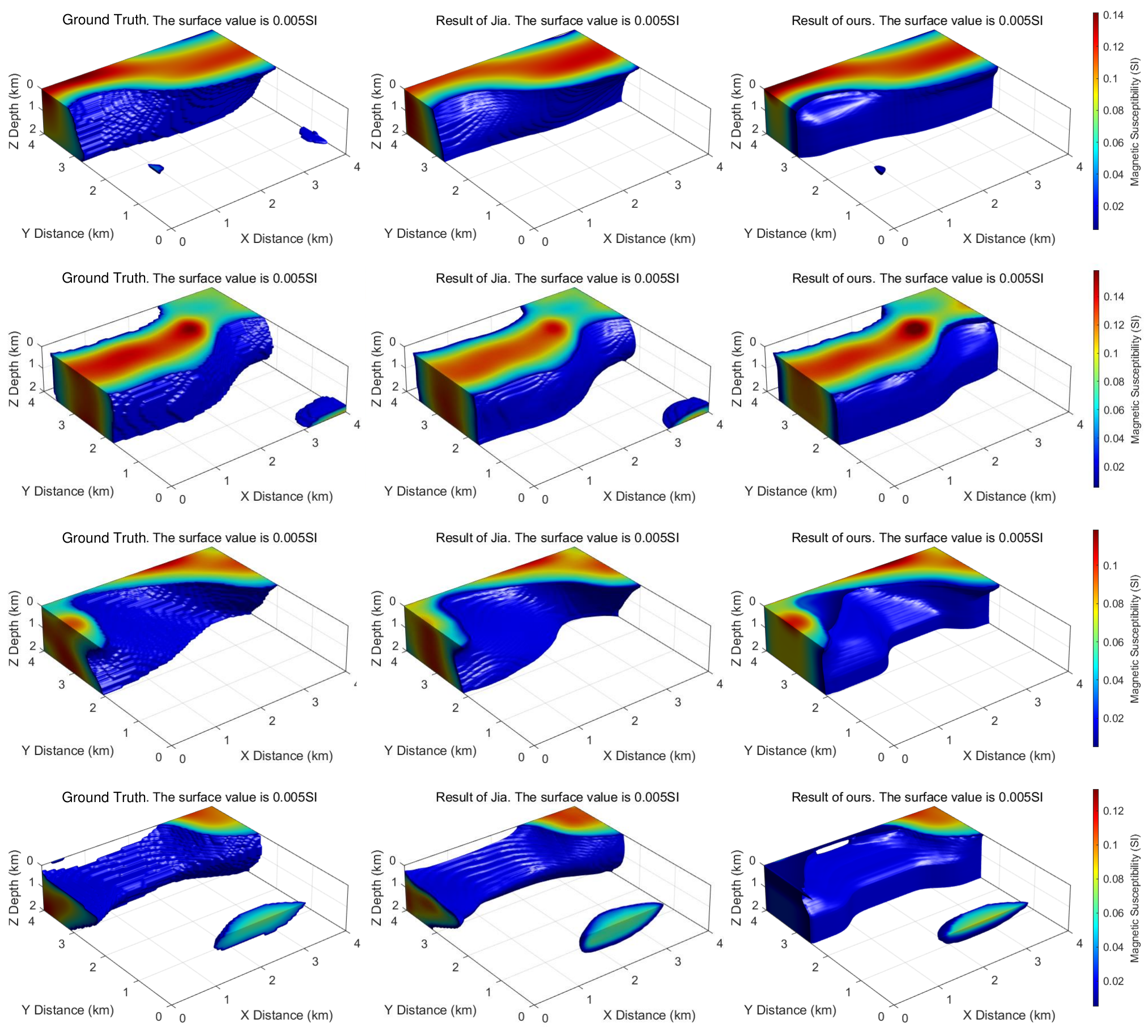}
\caption{Visual comparison of the estimated inversion results on synthetic data.}
\label{fig_synthetic}
\end{figure*}

\subsection{Closed Loop and Self-Supervised Learning.}

The intact ground truth (GT) of the subsurface magnetic susceptibility distribution is unavailable, hence there is no GT for magnetic inversion. The forward model devotes to obtaining 2D surface magnetic anomaly from 3D subsurface magnetic susceptibility distribution, which is the reverse operation of the inversion model. Thus, SSKMI builds two neural networks to realize a closed loop between inversion and forward. Since the parameters of the forward model are determined by Coulomb’s law, the parameters of the inversion model can be optimized by the closed loop. Specifically, the inversion model can be optimized by minimizing the mean absolute error (MSE) between the original and estimated surface magnetic anomalies.

Since magnetic inversion is an ill-posed problem, we proposed to constrain it by a guideline in the depth direction. Assuming that the magnetic susceptibility distribution is shaped in $[h,w,d]$, where $d$ stands for the depth. The guideline is 1D information whose length is $d$, which can be derived from field data, the expression is as follows:
\begin{equation}
L=\mathcal{G}(M),
\label{eq_line}
\end{equation}
where $\mathcal{G}$ denotes the function from the magnetic susceptibility distribution $M$ to the guideline $L$, which can be sampling or average operation.

\subsection{Objective Function and Optimization Strategy.}

The expression of the objective function of SSKMI is as follows:
\begin{equation}
\begin{aligned}
\min\ \lambda_1\ell_1(Z, Z')& + \lambda_2\ell_1(L,L') + \lambda_3||M'||_1 + \lambda_4TV(M'),\\
s.t.\ &\left\{
  \begin{aligned}
  M'=&\mathcal{I}(Z,L;\theta),\\
  Z'=&\mathcal{F}(M'),\\
  L'=&\mathcal{G}(M'),\\
  TV(M')=&||\triangledown M'||_2^2,
  \end{aligned}
\right.
\end{aligned}
\label{eq_loss}
\end{equation}
where $Z$ and $L$ stand for the observed magnetic anomaly and guideline, while $M'$, $Z'$, and $L'$ denote the estimated subsurface magnetic susceptibility distribution, the re-estimated magnetic anomaly, and the generated guideline, respectively. Meanwhile, $\lambda_i,i\in 1,2,3,4$ represent the weight parameters, while $\theta$ denotes the learnable parameters of the inversion model $\mathcal{I}$. The first and second parts of the objective function are the loss function of the main loop and the auxiliary loop. Besides, the final two parts prevent noise and fluctuation, where $TV$ stands for total variation \cite{rudin1992nonlinear}.

Adam \cite{kingma2015adam} is employed as the optimizer with a learning rate decay method. The training is started with $\lambda_1:\lambda_2:\lambda_3:\lambda_4=1:5:0.1:0.1$ in the first 400 iterations, by which the proposed SSKMI can quickly match the guideline and roughly determine the longitudinal distribution of the 3D susceptibility. Then, the proposed SSKMI is concerned with minimizing the difference between the observed and the estimated magnetic anomaly with $\lambda_1:\lambda_2:\lambda_3:\lambda_4=1:0.5:1:1$ in the next 2100 iterations. In the final 500 iterations, the weight parameters are set to $\lambda_1:\lambda_2:\lambda_3:\lambda_4=1:0.5:1:1$ to prevent over-fitting of SSKMI.

\section{Experimental Analysis }\label{Experiments}

\begin{figure}
\noindent\includegraphics[width=\linewidth]{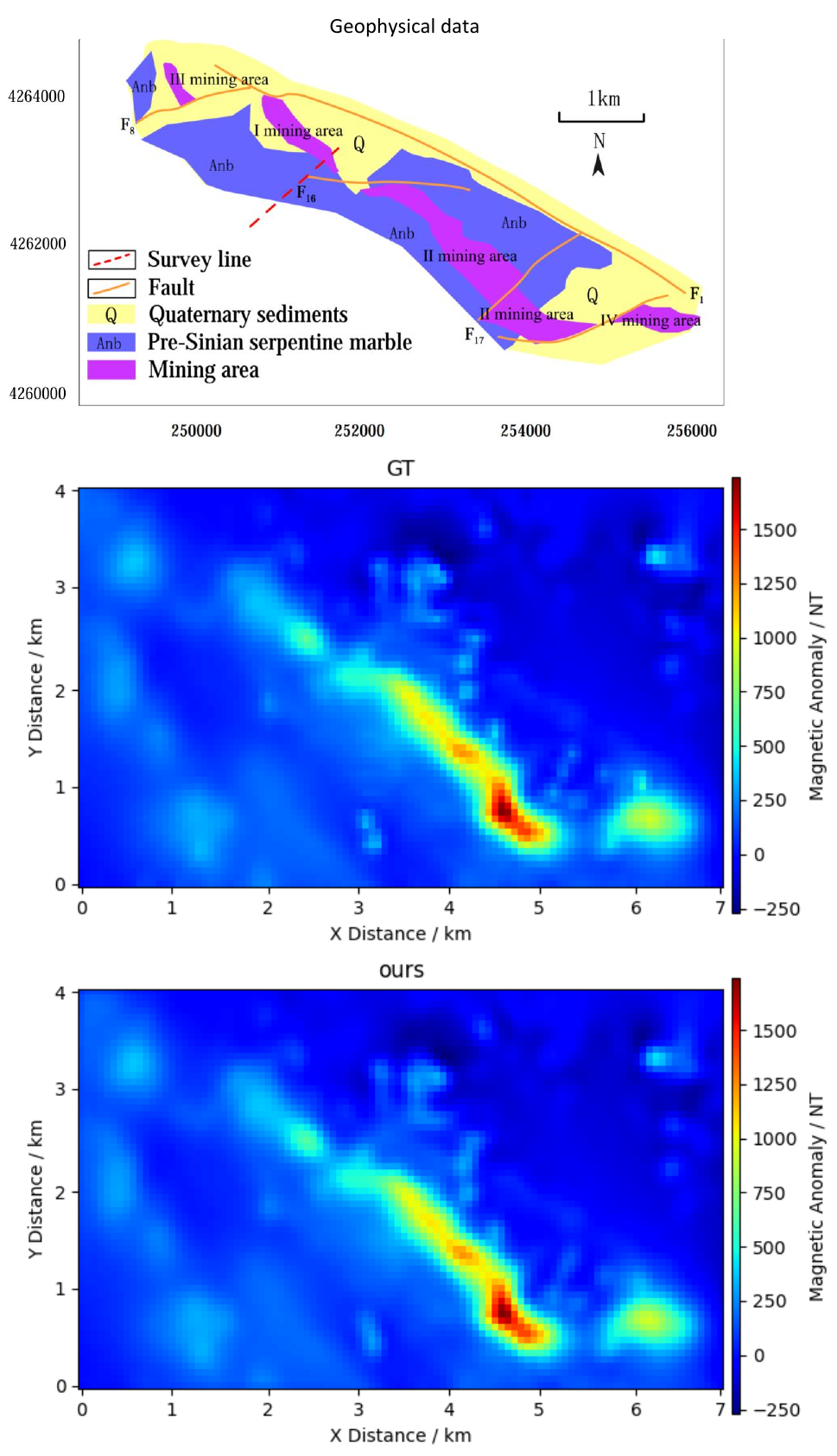}
\caption{Visual comparison of background geophysical data, the observed and re-estimated magnetic anomalies. ``GT'' is the observed magnetic anomaly. The re-estimated magnetic anomaly of our proposed method consists of the background geophysical data and the observed magnetic anomaly.}
\label{fig_2d}
\end{figure}


\begin{figure}[t]
\noindent\includegraphics[width=\linewidth]{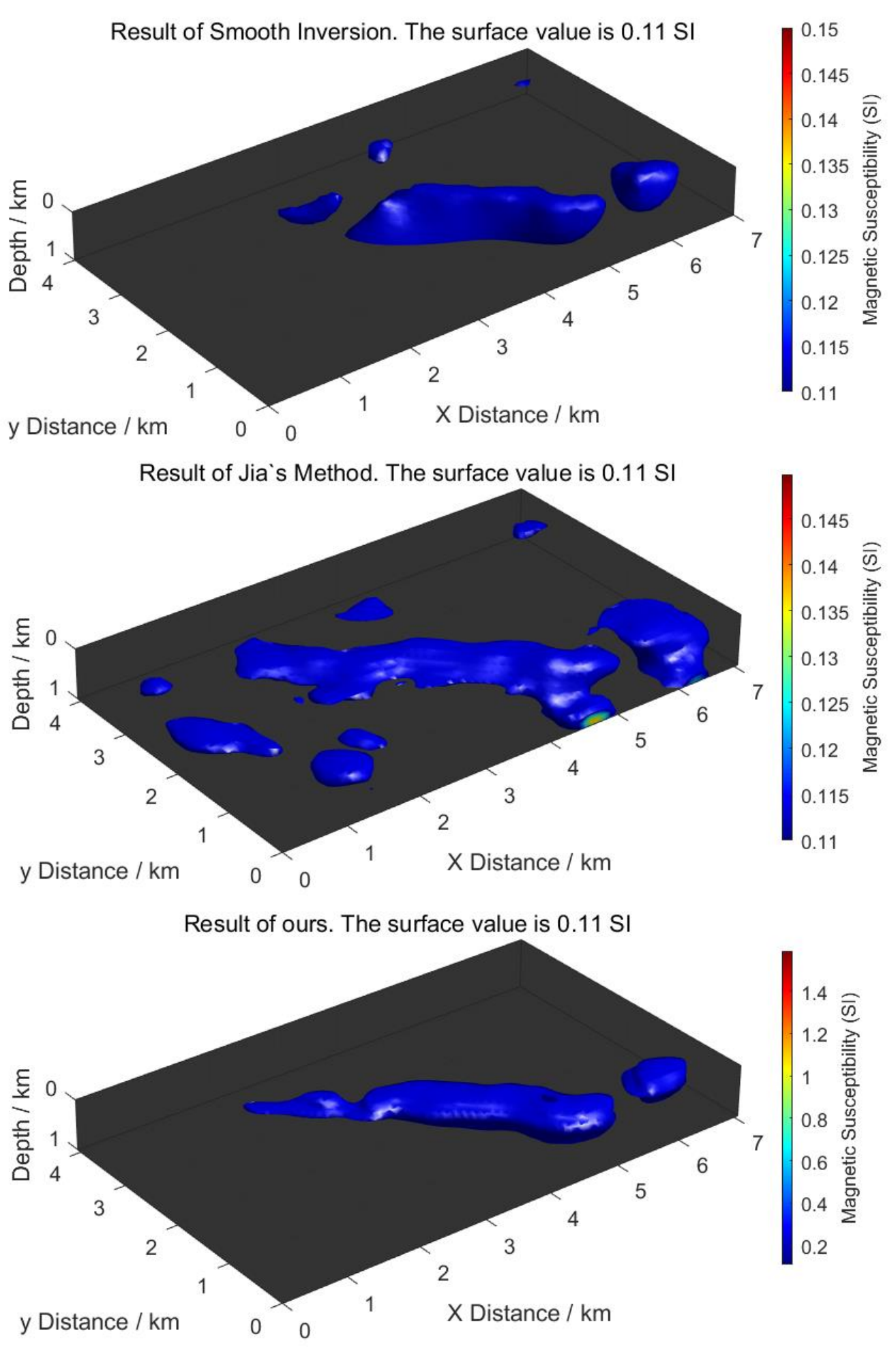}
\caption{Visual comparison of the estimated inversion results. The surface value is 0.11 SI. ``Smooth Inversion'' is the result of a traditional magnetic inversion method proposed by Gao et al. \cite{Gao2019}, while ``Jia'' denotes the SOTA DL-based open-loop end-to-end magnetic method proposed by Jia et al. \cite{jia2023deep}.}
\label{fig_3d}
\end{figure}

\begin{table*}[ht]
\caption{Quantitative evaluation on synthetic data. The performance of the proposed method with that of the contrast is compared in terms of mean absolute error(MAE), signal noise ratio (SNR), and cross-correlation (XCOR). The best are highlighted in \textcolor{red}{red}.}
\label{tab_syn}
\centering
\begin{tabular}{l l | c c c | c c c}
\hline
 \multirow{2}*{Method} & \multirow{2}*{Type} & \multicolumn{3}{c|}{Magnetic Anomaly} & \multicolumn{3}{c}{Susceptibility} \\
 & & MAE $\downarrow$ & SNR $\uparrow$ & XCOR $\uparrow$ & MAE $\downarrow$ & SNR $\uparrow$ & XCOR $\uparrow$ \\
\hline
 Jia \cite{jia2023deep} & Supervised & 0.0242 & 24.1484 & 0.9959 & \textcolor{red}{0.0002} & \textcolor{red}{15.5009} & \textcolor{red}{0.9780} \\
 ours & Self-Supervised & \textcolor{red}{0.0074} & \textcolor{red}{42.5738} & \textcolor{red}{0.9988} & 0.0008 & 7.0619 & 0.8242 \\
\hline
\end{tabular}
\end{table*}

The proposed method is evaluated on synthetic data and field data in this section.

\subsection{Evaluate on Synthetic Data}

Generally, DL-based magnetic inversion methods require high-quality training data. These methods are trained and evaluated on synthetic data and then applied to inference on field data.
Since the proposed method does not require extra training data, it can be evaluated on the test set of synthetic data \cite{jia2023deep} without training. Besides, Jia's method, the state-of-the-art (SOTA) magnetic inversion model based on supervised deep learning, is selected as the contrast method. 

The synthetic data set provided in Jia's paper \cite{jia2023deep} is utilized in this paper, which is randomly split into the training set and test set.
Jia's method is trained on the $8k$ paired training set and then evaluated on the $4k$ test set. Specifically, Jia's method is an end-to-end deep learning method and learns the map from the surface magnetic anomaly to the subsurface susceptibility. Differently, the proposed self-supervised method is trained and evaluated on the test set directly by the closed loop between the inversion and forward model.

Fig. \ref{fig_synthetic} visually compares the estimated inversion results on synthetic data. Since the randomly selected test set of the synthetic data is independently identically distributed with others (the training set), Jia's method achieved outstanding performance with suspicion of over-fitting. What is noteworthy is that the proposed self-supervised method has also obtained exceptional performance.

Quantitative evaluation on synthetic data is shown in Table \ref{tab_syn}. The performance of the proposed method with that of the contrast is compared in terms of mean absolute error(MAE, lower is better), signal noise ratio (SNR, higher is better), and cross-correlation (XCOR, higher is better). The supervised method obtains more similar estimate results of magnetic susceptibility, while the proposed self-supervised method achieves decent results. It's worth noting that the proposed self-supervised method has achieved brilliant performance in the reconstruction of magnetic anomalies. It demonstrates that the proposed self-supervised method has achieved feasible inversion results, which are not fully consistent with the answers in the data set.

\subsection{Evaluate on Field Data}

\begin{figure}
\centering
\noindent\includegraphics[width=\linewidth]{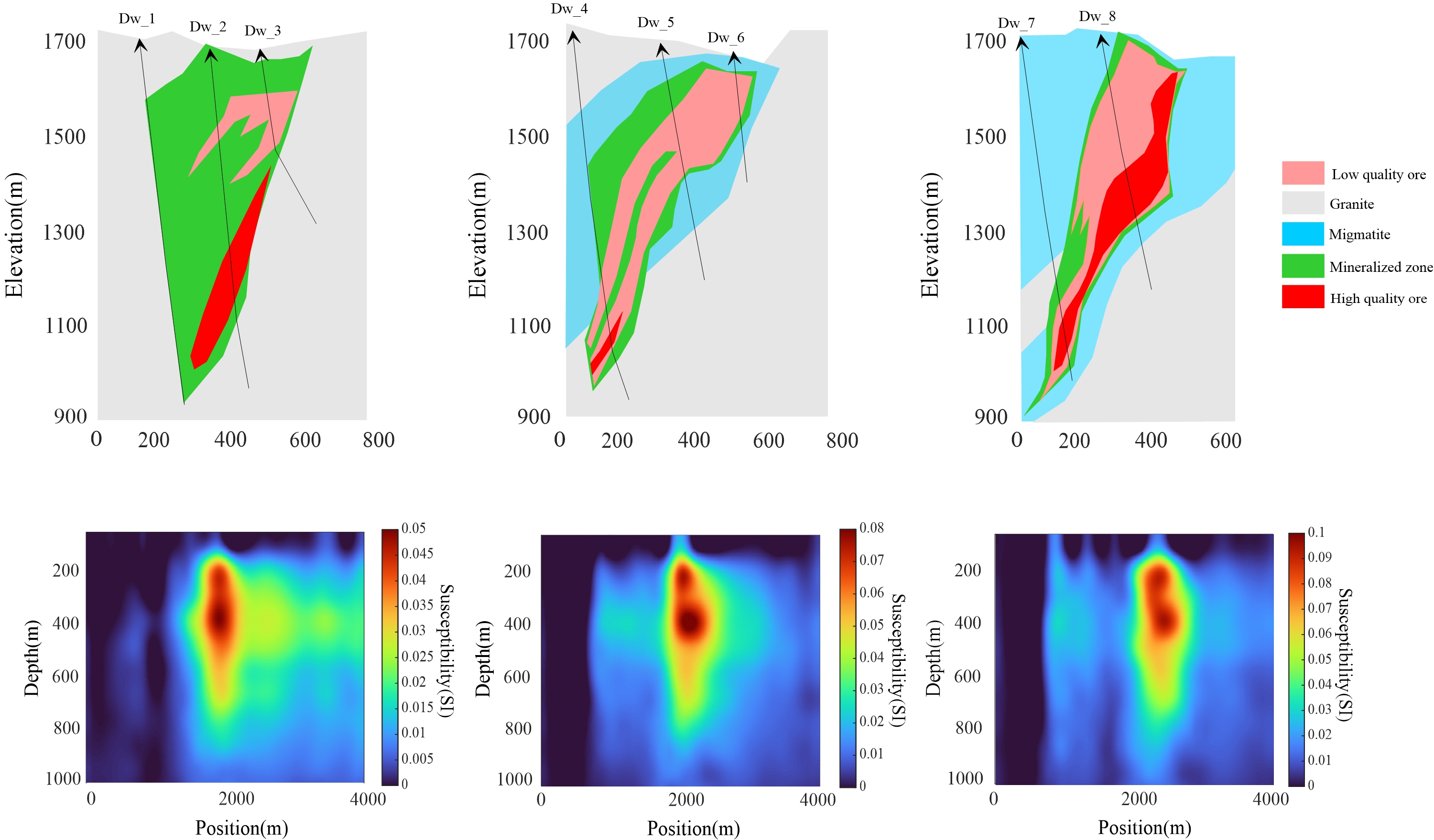}
\caption{Verify the magnetic inversion performance by ore deposit in the field data.}
\label{fig_ore}
\end{figure}

\begin{figure}
\centering
\noindent\includegraphics[width=\linewidth]{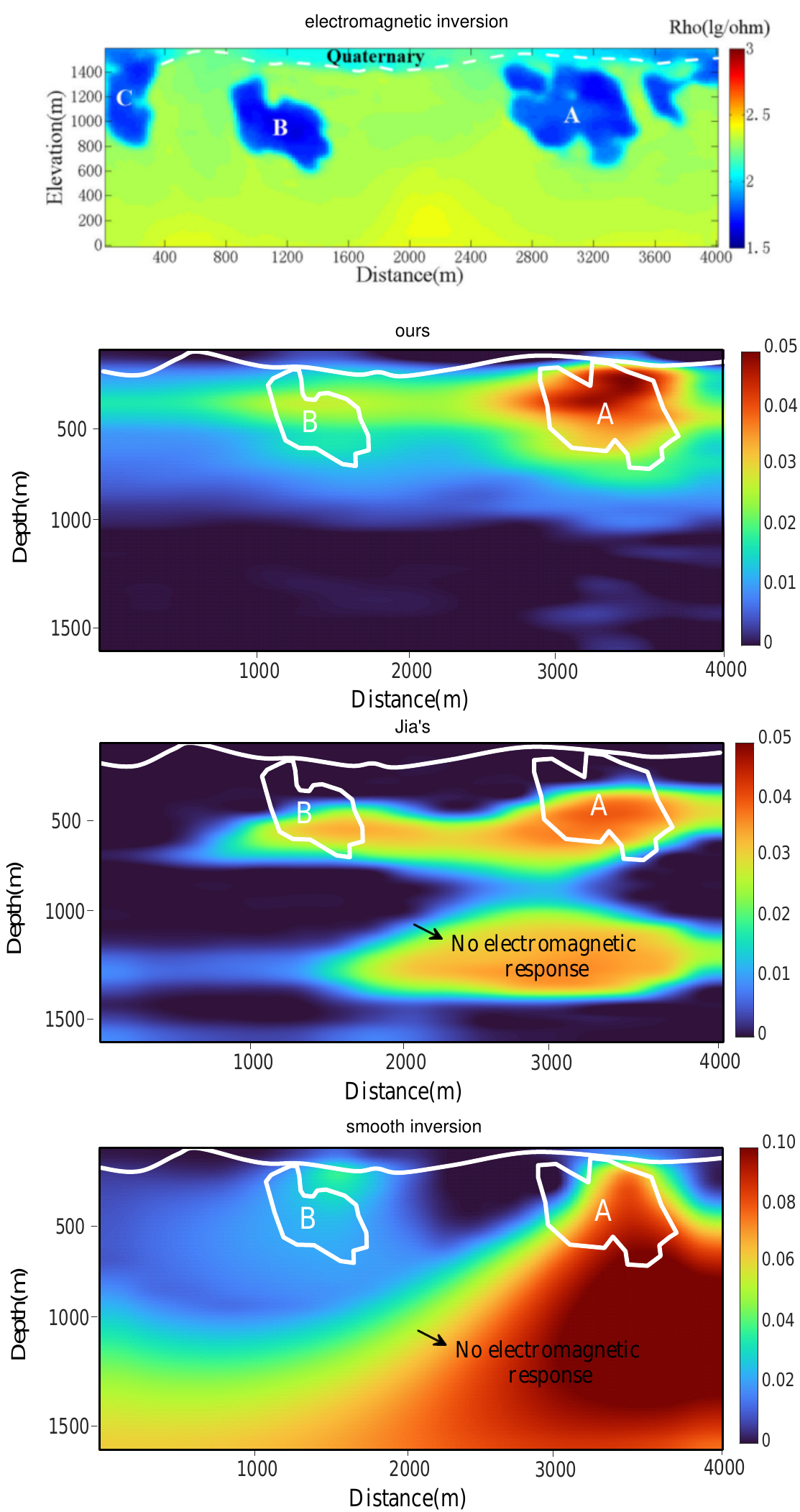}
\caption{Visual comparison of the estimated inversion results and electromagnetic inversion result of EMRNet \cite{jia2022emrnet} on the red dashed line in Fig. \ref{fig_2d}.}
\label{fig_ele}
\end{figure}

The proposed method is further evaluated on field data. The observed 2D surface magnetic anomaly in Fig. \ref{fig_2d} comes from the Jinchuan deposit, northwest China \cite{chai1992characteristics}.
The Jinchuan copper-nickel deposit in northwest China is 6500 meters long and 20 to 527 meters wide. The rock mass can be divided into four sections by the northeast thrust fault. Specifically, the Jinchuan deposit is divided into III, I, II, and IV zones from west to east \cite{cao2009process}. 
The eastern part of the body intrudes into the mixed rocks, the orientation of the body is at an angle of 5° to 10° with the surrounding rocks, with the upper part of the body cutting diagonally into the marble and the lower part cutting into the mixed stones.
The dip of the rock body's upper and lower plates is inconsistent. Specifically, the lower plate of the enclosing rock has a steep dip, while the upper plate of the rock has a slow dip. The direct surrounding rocks of the ultramafic body are mainly dacite, mixed rock, and gneiss. Besides, the ore-bearing ultramafic rock body is an irregularly shaped rock wall, covered by the quaternary sediments at the edge \cite{zhang2020deep}.

The ground truth of the subsurface 3D magnetic susceptibility distribution is unavailable. Thus, this paper evaluates the inversion results by reconstructing the magnetic anomaly, i.e. closed loop of inversion-forward. An average vector of the magnetization matrix in the depth direction generated based on well-log and expert knowledge is given as the guideline. As shown in Fig. \ref{fig_2d}, the re-estimated surface magnetic anomaly of the proposed method is similar to the observed surface magnetic anomaly in the field data.
Besides, the result obtained by the proposed method is rich in detail and consists of the background geological data.
Specifically, the proposed method has recovered four mining areas consistent with the background geophysical data.
Meanwhile, in the southwest direction, the magnetic susceptibility anomaly and the magnetic anomaly intensity have begun to take shape. The peak intensity of the anomalous individuals varied greatly, and the lowest one was 130nT. The highest anomaly zone peak was close to 600nT. The minimum length of an anomaly is about 500m. The variation of anomaly width is roughly related to anomaly intensity, and the variation range is $400\sim1000$m. The anomaly arrangement trend is consistent with the ore deposit trend, and the distribution form of the anomaly is exactly corresponding to the known ore belt. Some anomalies are connected with the ore deposit with a small distance and low-value anomaly.

Jia's method \cite{jia2023deep} is the SOTA magnetic inversion model based on deep learning, which is an end-to-end method selected as a contrast method. Besides, the smooth inversion method proposed by Gao et al. \cite{Gao2019}, a traditional 3D magnetic inversion method, is employed as another contrast method.
Fig. \ref{fig_3d} shows the visual comparison of the inversion results of the proposed method and the contrast methods. It can be seen that the deeper magnetic susceptibility estimated by Gao's method is similar to the near-surface magnetic susceptibility, and the resolution is low. Besides, the anomaly in $0\sim2$ km in the x-direction is inconspicuous in Gao's result, which is unsatisfactory. Meanwhile, although the methods proposed by Jia et al. can obtain the best performance on the synthetic data \cite{jia2023deep}, it fails to estimate a visually pleasing subsurface susceptibility on the field data. The performance of Jia's method on the field data is limited by the difference between the synthetic data and the field data.


Table \ref{tab_2d} shows the quantitative evaluation of field data. The performance of the proposed method (SSKMI) with that of the contrast is compared in terms of MAE, SNR, and XCOR. As shown in Table \ref{tab_2d}, the result generated by the proposed SSKMI is the best since it has obtained the smallest MAE and the biggest SNR and XCOR. Besides, the re-estimated result of Jia's method is better than that of the traditional inversion method, but its performance is limited by the difference between synthetic and field data. Since the proposed SSKMI is self-trained on the target field data based on professional knowledge, it can overcome the drawback of the deep learning methods in geophysics.

\begin{figure}
\centering
\noindent\includegraphics[width=\linewidth]{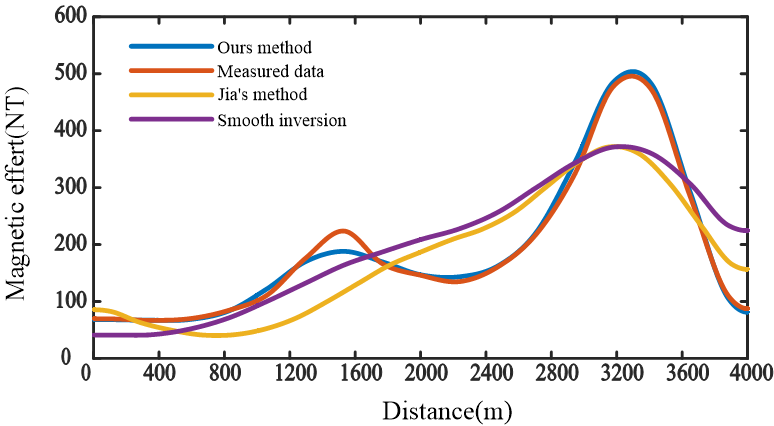}
\caption{Visual comparison of the estimated magnetic anomalies and measured data on the red dashed line in Fig. \ref{fig_2d}.}
\label{fig_line}
\end{figure}

\begin{figure}
\centering
\noindent\includegraphics[width=\linewidth]{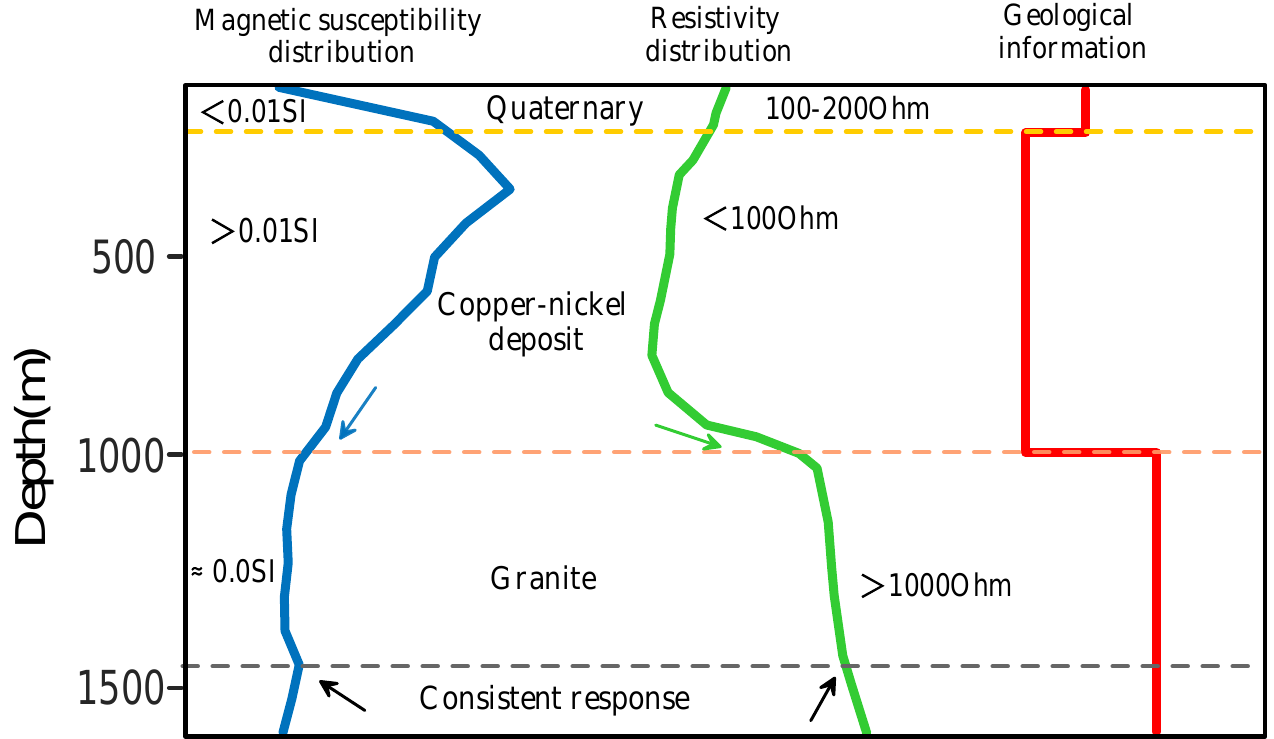}
\caption{Curve of depth geological information, resistivity information, and magnetic susceptibility information of anomaly A in Fig. \ref{fig_ele}.}
\label{fig_res}
\end{figure}

The performance of our proposed method is further evaluated by the ore deposit and electromagnetic inversion.
As shown in Fig. \ref{fig_ore}, the comparison results with the ore deposit verify the effectiveness of the proposed method. Specifically, the estimated 3D subsurface magnetic susceptibility distribution is consistent with the ore deposit in position and value. Besides, the extended depth of the inversion result can exceed 600m, effectively guiding the prospecting work. 

Fig. \ref{fig_ele} shows the visual comparison of the estimated inversion results and electromagnetic inversion result of EMRNet \cite{jia2022emrnet} on the red dashed line in Fig. \ref{fig_2d}. The magnetic inversion results in Fig. \ref{fig_ele} are provided by our proposed method, Jia's method \cite{jia2023deep}, and the smooth inversion method \cite{Gao2019}. The electromagnetic inversion result of EMRNet demonstrates that the Quaternary system is buried 100 meters below the surface. Anomaly A corresponds to the inversion of magnetic susceptibility, which may be related to the strong magnetic geological body. Besides, anomaly A indicates that there are two kinds of physical properties for ore-body prospects. The inversion results of anomaly B low-resistivity anomaly correspond to the three groups of inversion results. The development of low-resistivity anomaly may have a certain correlation with the strong magnetic geological body. The magnetic susceptibility of anomaly B location increases slowly, which may be beneficial to ore prospecting.

The profile in Fig. \ref{fig_ele} shows that the result of the proposed method consists of the electromagnetic inversion result in the Quaternary boundary and two strong magnetic geological bodies. Besides, the results of Jia's method and the smooth inversion method cannot match the anomalies in the electromagnetic inversion result. Meanwhile, both of the two control methods have provided low-reliability magnetic anomalies where there is no electromagnetic response.
Besides, a visual comparison of the estimated magnetic anomalies and measured magnetic anomaly on the red dashed line in Fig. \ref{fig_2d} is shown in Fig. \ref{fig_line}. It demonstrates that the re-estimated magnetic anomaly of our proposed method is the most similar to the measured magnetic anomaly. In conclusion, the proposed closed-loop inversion network is far superior to the open-loop network (Jia's method \cite{jia2023deep}) and traditional inversion (smooth inversion \cite{Gao2019}) in magnetic inversion.

Fig. \ref{fig_res} shows the curve of depth geological information, resistivity information, and magnetic susceptibility information of anomaly A in Fig. \ref{fig_ele}. Geological data and well-logging information can be used to estimate the depth relationship between the ore zone and the background rock mass. In anomaly A, the content of ore bodies is focused on 100 to 1000 meters.
Besides, within 100 meters are Quaternary sedimentary rocks, while over 1000 meters is full of mixed rock and marble.

\begin{table}[b]
\caption{Quantitative evaluation on field data. The performance of the proposed method with that of the contrast is compared in terms of mean absolute error (MAE), signal-to-noise ratio (SNR), and cross-correlation (XCOR). The best are highlighted in \textcolor{red}{red}.}
\label{tab_2d}
\centering
\begin{tabular}{l c c c }
\hline
    Method & MAE $\downarrow$ & SNR $\uparrow$ & XCOR $\uparrow$ \\
\hline
    Smooth \cite{Gao2019} & 0.1022 & 5.5511 & 0.8834 \\
    Jia \cite{jia2023deep} & 0.0711 & 8.1859 & 0.9282 \\
    ours & \textcolor{red}{0.0035} & \textcolor{red}{29.3839} & \textcolor{red}{0.9993} \\
\hline
\end{tabular}
\end{table}

\begin{figure}
\centering
\noindent\includegraphics[width=0.7\linewidth]{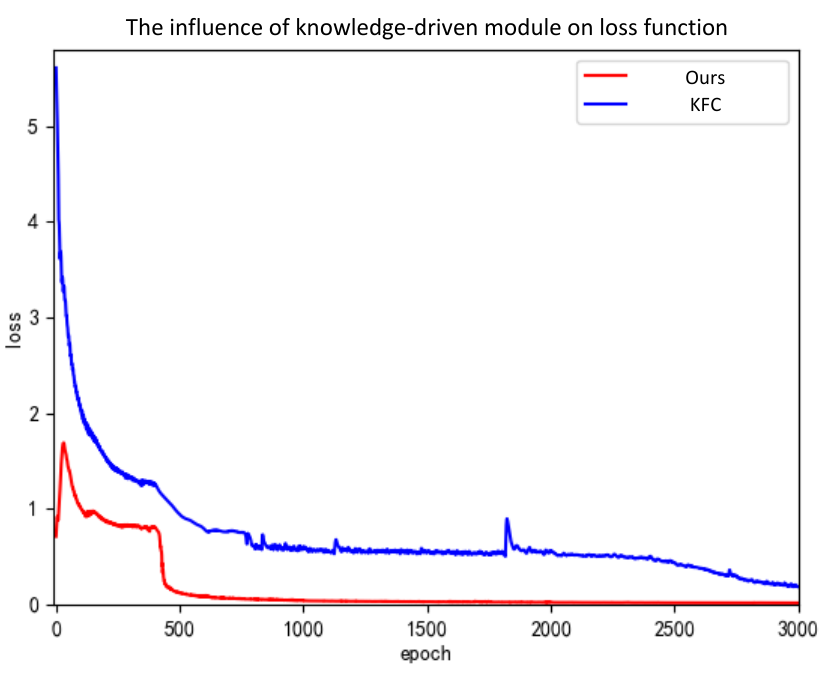}
\caption{The influence of knowledge-driven module on the training speed. MAE is selected as the loss function.}
\label{fig_loss}
\end{figure}

\begin{figure}[t]
\centering
\noindent\includegraphics[width=0.98\linewidth]{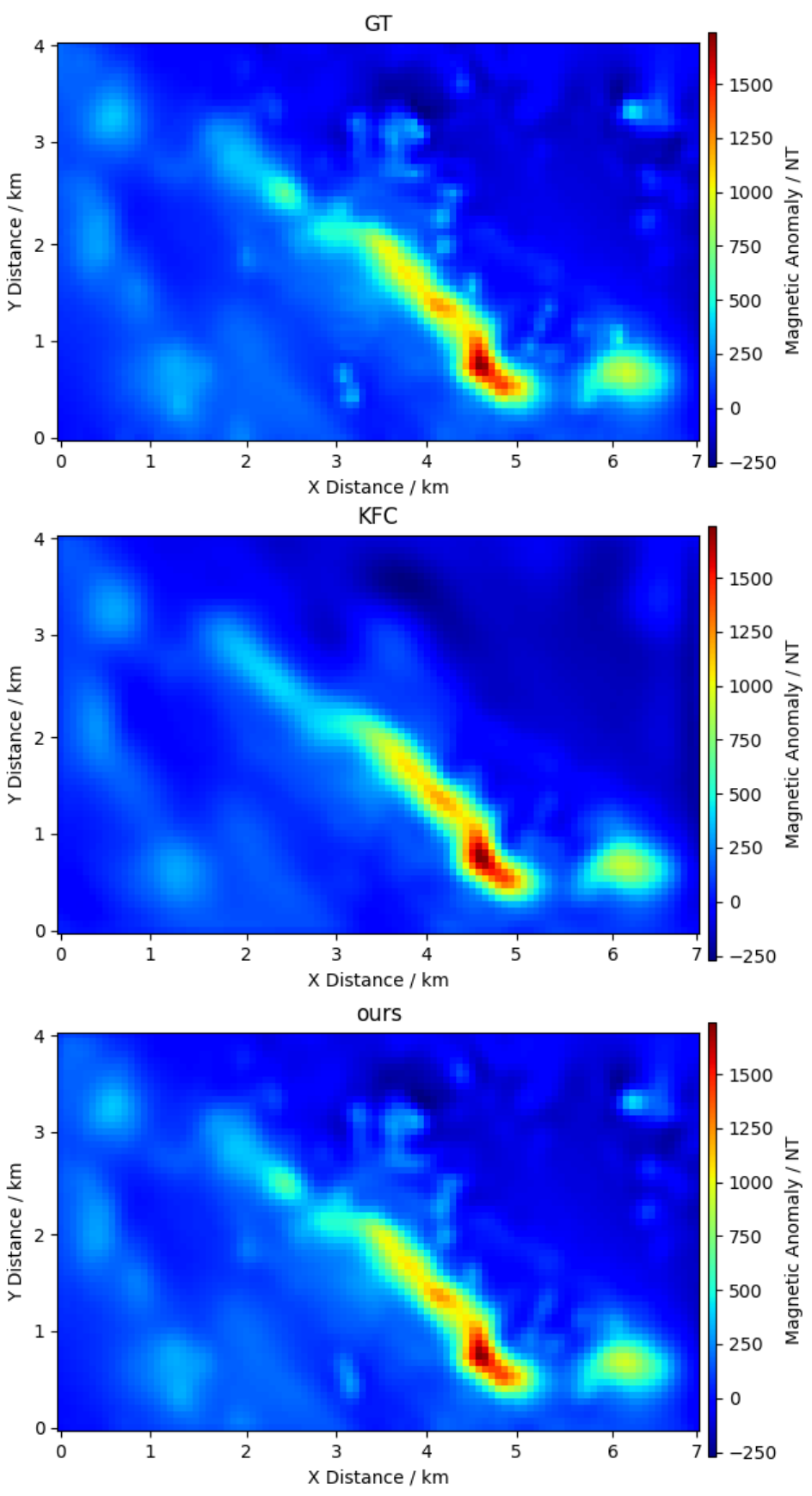}
\caption{Visual comparison of the re-estimated surface magnetic anomaly. ``GT'' stands for the ground truth (the observed magnetic anomaly), while KFC denotes to knowledge-free contrast.}
\label{fig_abl_2d}
\end{figure}

\begin{figure}[t]
\noindent\includegraphics[width=\linewidth]{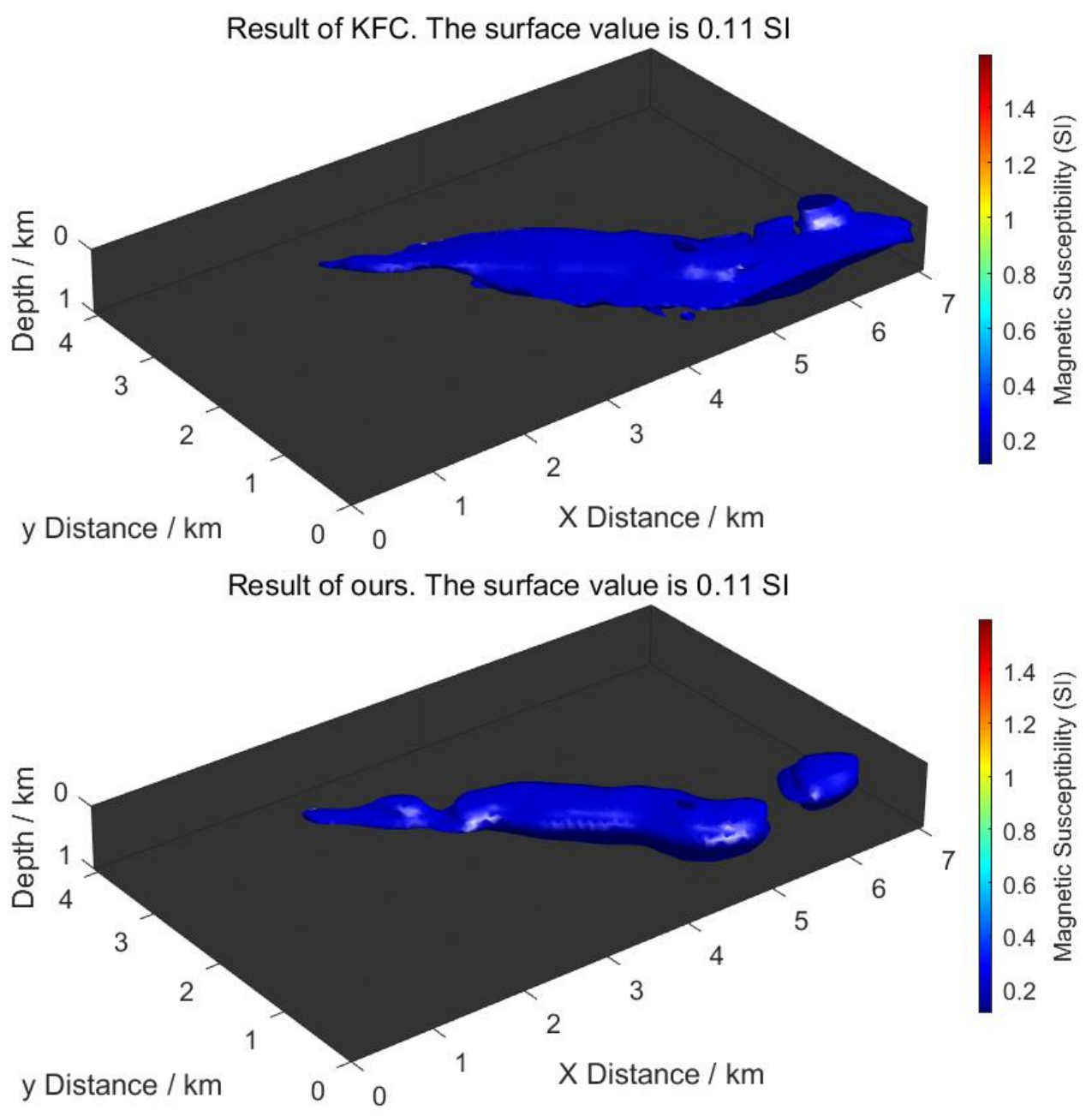}
\caption{Visual comparison of the inversion results. The surface value is 0.11 SI.}
\label{fig_abl_3d}
\end{figure}

\section{Ablation Study}

This paper explores the influence of the knowledge-driven module on training speed and performance in this section. An ablation study is designed to evaluate the influence of the knowledge-driven module. Specifically, a variant of SSKMI is utilized as the contrast method, which is the knowledge-free contrast (KFC).

This section first study on the influence of the knowledge-driven module on the training speed. It is necessary to state that $0\sim400$ iterations in training are utilized to fit the guideline and then the proposed method is concerned on fit the main closed loop. As shown in Fig. \ref{fig_loss}, the randomly initialized SSKMI has received an MAE less than $1$, and the value of MAE drops rapidly after $400$ iterations. Besides, the MAE of SSKMI dropped to a low enough state after about $1000$ iterations. On the contrary, when the knowledge-driven module is removed, the maximum MAE is more than $5$ and the MAE stagnates at $0.55$ in $1000\sim2000$ iterations. Experimental results have demonstrated the knowledge-driven module is indispensable for self-supervised DL magnetic inversion methods.

Then, the visual comparison of the re-estimated surface magnetic anomaly of SSKMI and KFC is shown in Fig. \ref{fig_abl_2d}.
As shown in Fig. \ref{fig_abl_2d}, the re-estimated surface magnetic anomaly of the proposed method is similar to the original surface magnetic anomaly in the field data. Besides, Table \ref{tab_abl_2d} shows the quantitative evaluation of the re-estimated magnetic anomaly in field data. The performance of the proposed method with that of the knowledge-free contrast (KFC) is compared in terms of MAE, SNR, and XCOR. As shown in Table \ref{tab_abl_2d}, the result generated by the proposed method is much better than that of KFC since it has obtained the smallest MAE and the biggest SNR and XCOR results.

Fig. \ref{fig_abl_3d} shows the visual comparison of the inversion results of the proposed method and KFC. It can be seen that KFC, which is free from the knowledge-driven module, failed to estimate the subsurface magnetic susceptibility. Besides, there is a deviation at the edge of KFC, which contradicts the observed surface magnetic data. Fig. \ref{fig_abl_3d} demonstrates that though KFC can reconstruct visually pleasing and high-accuracy surface magnetic anomaly, it cannot estimate a reliable subsurface susceptibility.
Since KFC fails to estimate the subsurface magnetic susceptibility, the knowledge-driven module is indispensable in the proposed self-supervised learning magnetic inversion method.

\section{Conclusion}\label{Conclusion}

To improve the magnetic inversion results, this paper proposed a self-supervised knowledge-driven 3D magnetic inversion method (SSKMI). The proposed SSKMI is based on the closed loop between the inversion and forward models. The inversion and forward models in the proposed SSKMI are realized by deep neural networks. Given that the forward model is based on Coulomb’s law and its parameters are fixed, SSKMI can optimize the parameters of the inversion model by the closed loop. Thus, the proposed method is based on deep learning but does not require additional training data. Besides, there is a knowledge-driven module in the proposed inversion model, which makes the DL-based inversion model explicable. Meanwhile, since magnetic inversion is an ill-posed task, this paper proposed to constrain the inversion model by a guideline in the auxiliary loop. The experimental results demonstrate that the proposed method is an effective DL-based magnetic inversion method.

\begin{table}[ht]
\caption{Quantitative evaluation of knowledge-driven module. The performance of the proposed method (SSKMI) with that of the knowledge-free contrast (KFC) is compared in terms of mean absolute error (MAE), signal-to-noise ratio (SNR), and cross-correlation (XCOR). The best are highlighted in \textcolor{red}{red}.}
\label{tab_abl_2d}
\centering
\begin{tabular}{l c c c }
\hline
    Method & MAE $\downarrow$ & SNR $\uparrow$ & XCOR $\uparrow$ \\
\hline
    KFC & 0.0477 & 10.2556 & 0.9471 \\
    ours & \textcolor{red}{0.0035} & \textcolor{red}{29.3839} & \textcolor{red}{0.9993} \\
\hline
\end{tabular}
\end{table}

\ifCLASSOPTIONcaptionsoff
  \newpage
\fi

\bibliographystyle{IEEEtran}
\bibliography{IEEEabrv,reference}
\end{document}